\documentstyle[preprint,aps]{revtex}
\begin{document}
\newcommand{\reell}{{\kern+.23em\sf{1}\kern-.61em\sf{1}\kern+.76em\kern-.25em}}
\title{Green's Function for a Neutron in Interaction with a Straight
Current Carrying Wire\thanks{This research was
supported in part by the Brazilian Research
Agencies CNPq and Finep.}}
\author{Rafael de Lima Rodrigues$^{a)}$ \\
Departamento de Ci\^encias Exatas
e da Natureza \\
Universidade Federal da Para\'{\i}ba \\
58.900-000 -- Cajazeiras, PB -- Brazil \\
Departamento de F\'\i sica
Universidade Federal da Para\'{\i}ba \\
58.100-000 -- Campina Grande, PB -- Brazil \\
and \\
Arvind Narayan Vaidya$^{b)}$ \\
Instituto de F\'{\i}sica \\
Universidade Federal do Rio de Janeiro \\
21.945-970 - Rio de Janeiro, RJ -- Brazil}
\maketitle

\begin{abstract}
We construct an integral representation for the momentum space
Green's function for
a Neutron in interaction with a straight current carrying wire.

\end{abstract}
\vspace{2cm}

\newpage

Consider an electrically neutral spin 1/2 particle of mass $M$ and
a magnetic moment $\mu\vec{\sigma}$ (a Neutron) in interaction
with an infinite straight wire carrying a current $I$ and located
along the $z$-axis. The magnetic field generated by the wire is given
by (we use units with $c=1=\hbar$)
\begin{equation}
\vec{B} = 2I\frac{(-y,x,0)}{(x^2+y^2)^{3/2}}
\end{equation}
where $x,y$ are cartesian coordinates in the plane perpendicular to
the wire. The Hamiltonian of the particle is given by
\begin{equation}
H = \frac{\vec{p}^{~\!2}}{2m} + \mu\vec{\sigma}.\vec{B}.
\end{equation}

The motion along the $z$-axis is free and will be ignored in the
following. Thus we get a two-dimensional problem with
\begin{equation}
H = \frac{\vec{p}^{~\!2}}{2m} +
2I\mu \frac{(-y\sigma_1+x\sigma_2)}{(x^2+y^2)^{3/2}}.
\end{equation}

The problem formulated above has a dynamical symmetry \cite{gp}.
{}From the form of the Hamiltonian there is one obvious constant of
motion
\begin{equation}
J_3 = x p_y-yp_x
 + \frac{\sigma_3}{2}.
\end{equation}
Also, the `Runge-Lenz vector' defined by
\begin{equation}
A_i = \frac{\varepsilon_{ij}x_j}{\sigma_1x_2-\sigma_2x_1} +
\frac{1}{4I\mu M} [p_i,J_3]_+,\quad (i=1, 2)
\end{equation}
satisfies the following commutation rules
\begin{equation}
[J_3,A_i] = i\ \varepsilon_{ij}\ A_j,\quad [A_i,H] = 0, \quad
[A_i,A_j] = -i\ H\ J_3 \ \frac{2M}{(2I\mu M)^2} .
\end{equation}

Thus for negative energy $(-E)$ eigenstates of $H$ we may put
\begin{equation}
J_i = A_i \left\{\frac{(2I\mu M)^2}{-2M\ E}\right\}^{1/2},\quad (i=1, 2)
\end{equation}
we get
\begin{equation}
[J_{\alpha},J_{\beta}] = i \ \varepsilon_{\alpha\beta\gamma}J_{\gamma}
\end{equation}
where $\alpha ,\beta = 1,2,3$. Thus $J_{\alpha}$ generate the algebra
of SO(3) in the space of negative energy eigenstates of $H$. For the
space of positive energy eigenstates of $H$ we change  $-E$ to
 $+E$
in above and get the commutation rules of the SO(2,1) algebra.

The SO(3) dynamical symmetry leads to an accidental degeneracy in the
bound state spectrum. One can show that $(E<0$)
\begin{equation}
\vec{A}^2 = 1+\frac{2M}{(2I\mu M)^2} \left(J_3^2+\frac{1}{4}\right)
\end{equation}
and
\begin{equation}
H = - \frac{(2I\mu M)^2}{2M(\vec{J}^2+\frac{1}{4})}
\end{equation}
so that the bound state spectrum is given by
\begin{equation}
E_{j + \frac{1}{2},M} = - \frac{(2I\mu M)^2}{2M(j+\frac{1}{2})^2}
\end{equation}
where $j=\frac{1}{2},\frac{3}{2}, \cdots$ and
$m=-j, -j+1, \cdots ,j$;
which exhibits the $(2j+1)$ fold degeneracy explicitly.


It may be mentioned that even if the bound state spectrum is
deduced on the basis of dynamical symmetry it is difficult to use it
for the construction of wave functions in the coordinate representation
which have an ackward form. As in the case of the Coulomb problem
the momentum representation allows us to use the dynamical symmetry
in an elegant manner.

As an example we formulate and solve the integral
equation for the Green's function $G(\vec{p},\vec{p}^{~\!\prime})$ in
momentum space. The calculation follows Schwinger`s \cite{j} technique
for the Coulomb Green's function except for one important modification.

To go over to the momentum representation we put
\begin{equation}
\vec{r} = i\ \vec{\bigtriangledown}_p
\end{equation}
and use
\begin{equation}
\bigtriangledown_p^2 \frac{1}{2\pi} \ln |\vec{p}-\vec{p}^{~\!\prime}| =
\delta (\vec{p}-\vec{p}^{~\!\prime})
\end{equation}
so that
\begin{equation}
\frac{i\vec{\bigtriangledown}_p}{\bigtriangledown_p^2} \Phi (\vec{p})
 = \frac{i}{2\pi} \int \ \Phi (\vec{p}^{~\!\prime})
\frac{(\vec{p}-\vec{p}^{~\!\prime})}{|\vec{p}-\vec{p}^{~\!\prime}|^2}
d^2p^{\prime}.
\end{equation}

The momentum space Green's function $G(\vec{p},\vec{p}^{~\!\prime})$ for energy
$E$ satisfies
\begin{equation}
(E-\frac{\vec{p}^{~\!2}}{2M}) G(\vec{p},\vec{p}^{~\!\prime}) \nonumber \\
- \frac{iI\mu}{\pi} \int d^2p^{\prime\prime}
\frac{\sigma_i\ \varepsilon_{ij}(\vec{p}-\vec{p}^{~\!\prime\prime} )_j}
{|\vec{p}-\vec{p}^{~\!\prime\prime}|^2}
G(\vec{p}^{~\!\prime\prime} ,\vec{p}^{~\!\prime}) =
\delta (\vec{p}-\vec{p}^{~\!\prime})
\end{equation}
where $\sigma_i (i=1,2,3)$ are the well known Pauli matrices and
$\varepsilon_{12}=1=-\varepsilon_{21}, \varepsilon_{ii}=0$.
Assuming that
\begin{equation}
E = - \frac{p^2_0}{2M}
\end{equation}
is real and negative $(p_0>0)$ we use the coordinates $(n_0,n_i)$ where
\begin{equation}
n_0 = \frac{p^2_0-\vec{p}^{~\!2}}{p^2_0+\vec{p}^{~\!2}} , \qquad n_i =
\frac{2p_0p_i}{p^2_0+\vec{p}^{~\!2}}
\end{equation}
and
\begin{equation}
n^2_0 + \vec{n}^{~\!2} = 1.
\end{equation}

It is convenient to treat $(n_0,\vec n)$ as a three dimensional vector
$n$. Here and in the following three dimensional vectors will be
written without the vector symbol. The scalar product of two
3-vectors $a$ and $b$ will be written as $a.b$, where $a=(a_0, a_1, a_2)$.

The element of area (solid angle) on the surface the unit sphere  of
equation (15) is
\begin{equation}
d\Omega = \frac{dn_1dn_2}{n_0} \left|_{|n|=1}\right. .
\end{equation}
Now
\begin{equation}
dn_1dn_2 = d^2p \quad J\left(\frac{p_i}{n_j}\right)
\end{equation}
where the Jacobian $J\left(\frac{p_i}{n_j}\right)$ is given by
\begin{equation}
J\left(\frac{p_i}{n_j}\right) =
\left(\frac{2p_0}{p_0^2+\vec{p}^{~\!2}}\right)^2 det (\reell\!\!-M)
\end{equation}
where the matrix elements of $M$ are
\begin{equation}
M_{ij} = \frac{2p_i\ p_j}{p^2_0+\vec{p}^{~\!2}}.
\end{equation}
Since for a matrix N
\begin{equation}
\ln \ det N=tr \ \ln N
\end{equation}
we get
\begin{equation}
\ln \ det(\reell\!\!-M) = \ln \ n_0
\end{equation}
where we use the property
\begin{equation}
M^2 = M(1-n_0).
\end{equation}
Thus
\begin{equation}
d\Omega  = \left(\frac{2p_0}{p^2_0+\vec{p}^{~\!2}}\right)^2 d^2p
\end{equation}
which gives
\begin{equation}
\delta (\Omega - \Omega ^{\prime}) =
\left(\frac{p^2_0+\vec{p}^{~\!2}}{2p_0}\right)^2 \delta
(\vec{p}-\vec{p}^{~\!\prime}).
\end{equation}
Further
\begin{eqnarray}
(n-n^{\prime})^2 &=& (n_0-n^{\prime}_0)^2 + (\vec{n}-\vec{n}^{\prime})^2
\nonumber \\
 &=&\frac{(1+n_0)(1+n^{\prime}_0)}{p^2_0} (\vec{p}-\vec{p}^{~\!\prime})^2 .
\end{eqnarray}
We define
\begin{equation}
\Gamma (\Omega, \Omega^{\prime}) =
- \frac{p_0^4}{M(1+n_0)^{3/2} (1+n^{\prime}_0)^{3/2}}
\hat{U}^{-1}(n)G(\vec{p},\vec{p}^{~\!\prime})\ \hat{U}(n^{\prime})
\end{equation}
and
\begin{equation}
D(\Omega, \Omega ^{\prime}) = -\frac{1}{4\pi} \hat{U}^{-1}(n) \
\frac{2i(\vec{\sigma}\wedge (\vec{p}-\vec{p}^{~\!\prime}))}{p_0}
\ \frac{(1+n_0)^{1/2}(1+n^{\prime}_0)^{\frac{1}{2}}}{(n-n^{\prime})^2} \
\hat{U}(n^{\prime})
\end{equation}
We get
\begin{equation}
\Gamma (\Omega , \Omega^{\prime}) - \nu \int D(\Omega , \Omega^{\prime\prime})
\Gamma (\Omega^{\prime\prime} , \Omega^{\prime}) d\Omega^{\prime\prime}  =
\delta (\Omega -\Omega^{\prime})
\end{equation}
where
\begin{equation}
\nu = \frac{2I\mu M}{p_0}
\end{equation}
which {\it looks} covariant under 3-dimensional rotations
due to the notational artifice.

The introduction of the matrices $\hat{U}(n)$ although permitted
would appear to be redundant. However, if we omit them the form of
$D(\Omega ,\Omega ^{\prime})$ would not display in a convenient fashion the
dependence on the 3-vectors $n$ and $n^{\prime}$, although
the left hand side of
equation (33) should do so. We determine $\hat{U}(n)$ by having
$D(\Omega , \Omega ^{\prime})$ satisfy a differential equation where its
rotational invariance is obvious.

We note that
\begin{equation}
\frac{\vec{\sigma}\wedge (\vec{p}-\vec{p}^{~\!\prime})}{(n-n^{\prime})^2 \
p^2_0}
(1+n_0)(1+n^{\prime}_0) = \frac{\vec{\sigma}\wedge
(\vec{p}-\vec{p}^{~\!\prime})}{|\vec{p}-
\vec{p}^{~\!\prime}|^2}.
\end{equation}
Also
\begin{eqnarray}
&& (\vec{\sigma}\wedge \vec{\bigtriangledown}_p)
\frac{(\vec{\sigma}\wedge (\vec{p}-\vec{p}^{~\!\prime})}{|\vec{p}-
\vec{p}^{~\!\prime}|^2} = 2\pi \ \delta (\vec{p}-\vec{p}^{~\!\prime}) \nonumber
\\
&& = 2\pi \left(\frac{(1+n_0)}{p_0}\right)^2 \ \delta (\Omega -
\Omega^{\prime}).
\end{eqnarray}
Thus
\begin{equation}
(\vec{\sigma} \wedge \vec{\bigtriangledown}_p) (1+n_0)^{\frac{1}{2}}
(1+n^{\prime}_0)^{\frac{1}{2}} \hat{U}(n) D\ \hat{U}^{-1}(n^{\prime}) =
- ip_0 \left(\frac{1+n_0}{p_0}\right)^2 \delta (\Omega - \Omega^{\prime}).
\end{equation}
Let
\begin{equation}
\hat{U}(n) = \sqrt{\frac{1+n_0}{2}} \hat{U}(n)
\end{equation}
then
\begin{equation}
\hat{U}^{-1}(n) \frac{ip_0}{1+n_0} (\vec{\sigma}\wedge
\vec{\bigtriangledown}_p) \hat{U}(n) D(\Omega , \Omega^{\prime}) =
\delta (\Omega - \Omega^{\prime}).
\end{equation}

The right hand side of equation (37) suggests the use of polar
coordinates, hence if the polar angles of $(n_0, \vec{n})$ are
$\theta$, $\phi$;
\begin{eqnarray}
&& p_1 = p_0 \tan \ \left(\frac{\theta}{2}\right) cos \phi, \nonumber \\
&& p_2 = p_0  \tan \ \left(\frac{\theta}{2}\right) sin \phi .
\end{eqnarray}
We them have
\begin{equation}
\frac{ip_0}{1+n_0} \vec{\sigma} \wedge \vec{\bigtriangledown}_p
= -i (\sigma .e_{\phi} \partial_\theta - \frac{\sigma
.e_p}{sin\theta} \partial_{\phi})
\end{equation}
where $e_p$ and  $e_{\phi}$ are unit vectors.

Thus the operator acting on $D(\Omega , \Omega^{\prime})$ in equation (39)
involves at most first order differential operators $\partial_{\theta}$
and $\partial_{\phi}$. If  it were invariant under rotations we can
make $D(\Omega , \Omega^{\prime})$ invariant too. The unique candidate in
that case is the operator
$
a+b \sigma .L
$
where $a$ and $b$ are constants. We choose $\hat{U}(n)$
so that this holds true.

Since
\begin{equation}
\sigma .L = -i (\sigma .e_{\phi} \partial_{\theta} + \sigma_3
\partial_{\phi} - \cot(\theta) +  \sigma .e_p \ \partial_{\phi})
\end{equation}
the choice
\begin{equation}
\hat{U}(n) = f(\theta ,\phi ) \mbox{exp} \ \left(\frac{ig(\theta ,\phi )}{2}
\sigma .e_{\phi}\right)
\end{equation}
maintains the - $i\sigma .e_{\phi} \partial_{\theta}$ term (apart
from other addicional terms without derivatives).

Working out the derivatives it is easy to see that one must choose
$g(\theta , \phi )=\theta$ which gives
\begin{eqnarray}
&& \hat{U}^{-1} \left(-i \vec{\sigma}.\vec{e}_{\phi}\partial_{\phi} +
i \frac{\vec{\sigma}.\vec{e}_p}{sin\theta} \partial_{\phi}\right) \hat{U}
\nonumber \\
&& = \sigma . L - i \hat{U}^{-1} \ \sigma .e_{\phi} (\partial_{\theta}\hat{U})
+ i \ \hat{U}^{-1} \ \frac{\sigma .e_p}{sin\theta} (\partial_{\phi} \hat{U})
\end{eqnarray}
where in the bracketed terms the derivatives act only on $\hat{U}$.

If we impose the condition that the last two terms are independent
of the $\sigma$-matrices we get
\begin{equation}
\hat{U}(n) = \frac{1+n_0}{2} + \frac{i}{2} sin \theta \
\sigma .e_{\phi}
\end{equation}
and then
\begin{equation}
\hat{U}^{-1}(n)  \frac{ip_0}{1+n_0} \vec{\sigma} \wedge
\vec{\bigtriangledown}_p \hat{U}(n) = \reell + \sigma .L.
\end{equation}
Hence
\begin{equation}
(\reell + \sigma . L) D(\Omega , \Omega^{\prime}) = \delta (\Omega ,
\Omega^{\prime}).
\end{equation}
Our choice of $\hat{U}(n)$ gives
\begin{equation}
\hat{U}(n) = \sqrt{\frac{1+n_0}{2}} \left(1+\frac{\sigma_3
\vec{\sigma}.\vec{p}}{p_0}\right)
\end{equation}
which allows us to calculate $D(\Omega , \Omega^{\prime})$ explicitly by
using equation (32). We get
\begin{equation}
D(\Omega , \Omega^{\prime}) = \frac{1}{2\pi} \frac{(1-\sigma .n \ \sigma
.n^{\prime})}{(n-n^{\prime})^2} .
\end{equation}

Although we know $D(\Omega , \Omega^{\prime})$ in a closed form it is
convenient to rewrite it in terms of the eigenfunctions of the
operator $\reell +\sigma .L$. The eigenfunctions of the operators
$J^2, L^2 , S^2\  \mbox{and}  \ J_3$
where
$S=\frac{\sigma}{2}$
and
$J = L+S$
are known to be $Y^{\pm}_{jm} (\theta ,\phi )$ where $\pm$
correspond to $j=\ell \pm \frac{1}{2}$  respectively \cite{BD}. Also
\begin{equation}
(\reell + \sigma .L) Y^{\pm}_{jm} (\theta ,\phi ) =
\pm (j+\frac{1}{2}) Y^{\pm}_{jm} (\theta ,\phi )
\end{equation}
so that
\begin{equation}
D(\Omega , \Omega^{\prime}) = \sum_{jm} \frac{1}{j+\frac{1}{2}}
\left\{
Y^{+}_{jm} (\theta ,\phi ) \
Y^{+^{\dagger}}_{jm}(\theta^{\prime},\phi^{\prime})
-Y^{-}_{jm}(\theta , \phi )
\ Y^{-^{\dagger}}_{jm} (\theta^{\prime}, \phi^{\prime})
\right\}.
\end{equation}
Going back to equation (33) we get
\begin{equation}
\Gamma (\Omega , \Omega^{\prime}) =
\sum_{jm} \left\{\frac{Y^{+}_{jm} (\theta ,\phi
)Y^{+^{\dagger}}_{jm}(\theta^{\prime},\phi^{\prime})}{1-\frac{\nu}{j+\frac{1}{2}}} +
\frac{Y^{-}_{jm} (\theta ,\phi ) Y^{-^{\dagger}}_{jm}(\theta^{\prime} , \phi^{
\prime})}{{1+\frac{\nu}{j+\frac{1}{2}}}}\right\}
\end{equation}
which for $\nu >0$ has poles at
\begin{equation}
\nu = \frac{2IM\mu}{p_0}  = j+\frac{1}{2}.
\end{equation}
This gives the bound state energies which are same as before.
The unnormalized bound state wave functions are
$\frac{1} {(1+n_0)^{3/2}}
\hat{U}^{-1}Y^{+}_{jm} (\theta ,\phi )$.

Next, one can do the sum indicated in equation (52) to write an
integral representation for $\Gamma (\Omega , \Omega^{\prime})$ which in
turn leads to one for the Green's function $G(\vec{p},\vec{p}^{~\!\prime})$.

This is done by using the identities:
\begin{equation}
 \frac{1}{1\mp \frac{\nu}{j+\frac{1}{2}}} = \frac{j+\frac{1}{2}}
{j+\frac{1}{2}\mp \ \nu}
= 1 \pm \frac{\nu}{j+\frac{1}{2} \mp \nu}
= 1 \pm \frac{\nu}{j+\frac{1}{2}} +
\frac{\nu^2}{(j+\frac{1}{2})(j+\frac{1}{2}\mp \nu)}
\end{equation}
and using
\begin{equation}
\frac{1}{n} = \int_0^1 \rho^{n-1} d\rho
\end{equation}
valid for $n>1$.
Although each form of the identity leads to a different looking integral
representation they are related through integration by parts. We
concentrate on the second form.

We have
\begin{equation}
\Gamma (\Omega ,\Omega^{\prime}) = \delta (\Omega ,\Omega^{\prime}) +
\nu\int^1_0 (\rho^{-\nu} \Delta^+ - \rho^\nu \Delta^-) d\rho
\end{equation}
where
\begin{equation}
\Delta^{\pm} = \sum_{jm} \rho^{j-\frac{1}{2}}\ Y^{\pm}_{jm}(\theta
,\phi ) \ Y^{\pm^{\dagger}}_{jm} (\theta^{\prime},\phi^{\prime}).
\end{equation}
Using the explicit forms of $Y^{\pm}_{jm}(\theta ,\phi )$ one can
show that
\begin{eqnarray}
&& \Delta^{+} = \frac{1}{4\pi} (\partial_{\rho} \rho +
\sigma .L) \Phi, \nonumber \\
&& \Delta^{-} = \frac{1}{4\pi}
\left(-\frac{1}{\rho}\sigma .L + \partial_{\rho}\right) \Phi
\end{eqnarray}
where
\begin{equation}
\Phi = \left((1-\rho )^2 + \rho (n-n^{\prime})^2\right)^{-\frac{1}{2}}.
\end{equation}
Calculating the derivatives we get
\begin{eqnarray}
&& \Delta^+ = \frac{1}{4\pi} (1-\rho \ \sigma .n \
\sigma .n^{\prime})\Phi^3  ,\nonumber \\
&& \Delta^- = \frac{1}{4\pi} (\rho - \sigma .n \
\sigma .n^{\prime})\Phi^3 .
\end{eqnarray}
Thus
\begin{equation}
 \Gamma (\Omega , \Omega^{\prime}) = \delta (\Omega , \Omega^{\prime})
+ \frac{\nu}{4\pi} \int^1_0 d\rho
\left(
\begin{array}{l}
\rho^{-\nu} (1-\rho
\sigma .n \ \sigma .n^{\prime})
-\rho^{\nu}(\sigma.n \ \sigma .n^{\prime}-\rho)
\end{array}
\right).
\end{equation}
Substituting this in equation (31) we get the integral representation
for $G(\vec{p},\vec{p}^{~\!\prime})$:
\begin{eqnarray}
&& G(\vec{p},\vec{p}^{~\!\prime}) = \frac{\delta (\vec{p}-
\vec{p}^{~\!\prime})}{E-T}
 + \frac{\nu}{8\pi M}\ \frac{1}{E-T} \ \frac{1}{E-T^{\prime}} \
\frac{1}{|\vec{p}-\vec{p}^{~\!\prime}|^2}
\nonumber \\
&&\left\{
\left(2M(E-\frac{T+T^{\prime}}{2}) - i\sigma_3 (\vec{p}\wedge
\vec{p}^{~\!\prime})
+ \frac{1}{2} |\vec{p}-\vec{p}^{~\!\prime}|^2\right)\right. \nonumber \\
&& (I(-\nu)-I(\nu) +I(1+\nu) -I(1-\nu))
\nonumber \\
&&\left. +ip_0 \ \vec{\sigma}\wedge (\vec{p}-\vec{p}^{~\!\prime})
(I(\nu) + I(-\nu) + I(1+\nu) + I(1-\nu)
)\right\} \nonumber
\end{eqnarray}
where $p_0$ and $\nu$ are given by equations (20) and (34). Also,
\begin{equation}
I(\nu) = \int^1_0 d\rho \frac{\rho^\nu (n-n^{\prime})^2}{\left\{(1-\rho)^2+\rho
(n-n^{\prime})^2\right\}^{\frac{3}{2}}},
\end{equation}

\begin{equation}
T = \frac{p^2}{2M} ,\qquad  T^{\prime} = \frac{p^{\prime}{}^2}{2M}
\end{equation}
and
\begin{equation}
(n-n^{\prime})^2 = -\frac{2E}{M}
\frac{|\vec{p}-\vec{p}^{~\!\prime}|^2}{(E-T)(E-T^{\prime})}.
\end{equation}

In conclusion we have obtained an integral representation for the
Green's function for a Neutron in interaction with a linear current.
The results were deduced for the case $E<0$. By analytic continuation
to $E>0$ one can identify the scattering amplitude if the distorted
free propagators can be identified. In contrast to the
Coulomb problem it is not easy to solve the scattering problem in the
coordinate representation by separation of variables. Although a
general momentum transfer dependence of $(\vec{p}-\vec{p}^{~\!\prime})^{-2}$ is
indicated many details need to be worked out. This will be done elsewhere.


\newpage
\begin{thebibliography}{99}
\bibitem[a)]{byline } E-mail: CENDFI76@BRUFPB (BITNET)

\bibitem[b)]{ byline} E-mail: IFT10034@UFRJ (BITNET)

\bibitem{gp}  G. P. Pron'ko and  Yu G. Stroganov, {\it Sov. Phys.} JETP
{\bf45} (1977) 1073.
\bibitem{r}  R. Bl\"{u}mel and  K. Dietrich,
{\it Phys. Rev.} {\bf A 43} (1991) 22.
\bibitem{i}  I. Voronin, {\it Phys.  Rev.} {\bf A 43} (1991) 29.
\bibitem{j}  J. Schwinger,  {\it J. Math. Phys.} {\bf 5} (1964) 1607.
\bibitem{BD} J. D. Bjorken and S. D. Drell,
\it{ Relativistic Quantum Mechanics} (New York, McGraw Hill, 1965).


\end{thebibliography}
\end{document}